\documentclass[fleqn,usenatbib]{mnras}

\usepackage{newtxtext,newtxmath}
\usepackage[T1]{fontenc}

\DeclareRobustCommand{\VAN}[3]{#2}
\let\VANthebibliography\thebibliography
\def\thebibliography{\DeclareRobustCommand{\VAN}[3]{##3}\VANthebibliography}

\usepackage{graphicx}	% Including figure files
\usepackage{amsmath}	% Advanced maths commands

\newcommand{\kms}{\mbox{$\mathrm{km\,s^{-1}}$}}
\newcommand{\MSUN}{\mbox{$\mathrm{M_{\odot}}$}}
\newcommand{\MJUP}{\mbox{$\mathrm{M_\mathrm{Jup}}$}}
\newcommand{\RSUN}{\mbox{$\mathrm{R_{\odot}}$}}

\title[A dead cataclysmic variable?]{ZTF\,J021804.16+071152.93: a dead cataclysmic variable and potential solution to the missing period bouncers}

\author[S.~G.~Parsons et al.]{S.~G.~Parsons,$^{1}$\thanks{E-mail: s.g.parsons@sheffield.ac.uk}
A.~J.~Brown,$^{2}$
S.~L.~Casewell,$^{3}$
S.~P.~Littlefair,$^{1}$
J.~van Roestel,$^{4,5}$
A.~Rebassa-Mansergas,$^{6,7}$ \newauthor
R.~Murillo-Ojeda,$^{8}$
M.~Zorotovic,$^{9}$
M.~R.~Schreiber,$^{10}$
S.~Bagnulo,$^{11}$
M.~A.~Stroet,$^{11}$
N.~Castro Segura,$^{12}$\newauthor
V.~S.~Dhillon,$^{1,13}$
M.~J.~Dyer,$^{1,14}$ 
J.~A.~Garbutt,$^{1}$ 
M.~J.~Green,$^{15,16}$
D.~Jarvis,$^{1}$ 
M.~R.~Kennedy,$^{17}$
P.~Kerry,$^{1}$\newauthor
J.~McCormac,$^{12}$
J.~Munday,$^{12}$
I.~Pelisoli,$^{12}$
E.~Pike,$^{1}$
D.~I.~Sahman$^{1}$
and A. Yates$^{1}$
\\
$^{1}$Astrophysics Research Cluster, School of Mathematical and Physical Sciences, University of Sheffield, Sheffield S3 7RH, UK\\
$^{2}$ Hamburger Sternwarte, University of Hamburg, Gojenbergsweg 112, 21029 Hamburg, Germany \\
$^{3}$School of Physics and Astronomy, University of Leicester, University Road, Leicester LE1 7RH, UK\\
$^{4}$Institute of Science and Technology Austria, Am Campus 1, Klosterneuburg, 3400, Austria\\
$^{5}$Anton Pannekoek Institute for Astronomy, University of Amsterdam, 1090 GE, Amsterdam, The Netherlands\\
$^{6}$Departament de F{\'i}sica, Universitat Polit{\`e}cnica de Catalunya, c/Esteve Terrades 5, 08860 Castelldefels, Spain\\
$^{7}$Institut d'Estudis Espacials de Catalunya, c/ Esteve Terradas, 1, Edifici RDIT, Campus PMT-UPC, 08860 Castelldefels, Spain\\
$^{8}$Centro de Astrobiolog{\'i}a (CAB), CSIC-INTA, Camino Bajo del Castillo s/n, campus ESAC, 28692 Villanueva de la Ca{\~n}ada, Madrid, Spain. \\
$^{9}$Instituto de F{\'i}sica y Astronom{\'i}a, Universidad de Valpara{\'i}so, Av. Gran Breta{\~n}a 1111, Valpara{\'i}so, Chile\\
$^{10}$Departamento de F\'isica, Universidad T\'ecnica Federico Santa Mar\'ia, Av. Espa\~na 1680, Valpara\'iso, Chile\\
$^{11}$Armagh Observatory and Planetarium, College Hill, Armagh BT61 9DG, Northern Ireland, UK\\
$^{12}$Department of Physics, University of Warwick, Gibbet Hill Road, Coventry, CV4 7AL, UK\\
$^{13}$Instituto de Astrofisica de Canarias, E38205 La Laguna, Tenerife, Spain\\
$^{14}$Research Software Engineering, University of Sheffield, Sheffield, S1 4DP, UK\\
$^{15}$Homer L. Dodge Department of Physics and Astronomy, University of Oklahoma, 440 W. Brooks Street, Norman, OK 73019, USA\\
$^{16}$JILA, University of Colorado and National Institute of Standards and Technology, 440 UCB, Boulder, CO 80309-0440, USA\\
$^{17}$School of Physics, University College Cork, Cork, T12 K8AF, Ireland
}

\date{Accepted 2026 March 12. Received 2026 February 24}

\pubyear{2026}

\begin{document}
\label{firstpage}
\pagerange{\pageref{firstpage}--\pageref{lastpage}}
\maketitle

% Abstract of the paper
\begin{abstract}
It is predicted that half or more of all cataclysmic variables (CVs) should have evolved past the period minimum and now exist as so-called ``period bouncers'' where a white dwarf should be accreting from a Roche-lobe filling substellar companion. However, this prediction stands in stark contrast to observations, where only a few per cent of CVs are found in this evolutionary phase. A potential solution to this discrepancy is that a magnetic field emerges from within the white dwarf after the system has reached the period minimum. The transfer of angular momentum from the spin of the white dwarf into the orbit then pushes the two stars apart, detaching them for potentially billions of years. Here we present the discovery of ZTF\,J021804.16+071152.93, a detached $0.69\pm0.01$\,{\MSUN}, 19\,MG magnetic white dwarf plus $37\pm5$\,{\MJUP} brown dwarf binary with an orbital period of 1.7 hours. The kinematics of the system indicate that it is a high probability member of the galactic thick disk. However, this strongly disagrees with the much younger age of the system obtained from the white dwarf parameters, implying that the system may have been accreting in the past. This system is therefore consistent with having detached as a result of the emergence of the magnetic field of the white dwarf when the system was still mass transferring, and may represent the ultimate fate for many (perhaps even most) CVs.
 
\end{abstract}

\begin{keywords}
binaries: eclipsing -- stars: fundamental parameters -- white dwarfs -- brown dwarfs
\end{keywords}

%%%%%%%%%%%%%%%%%%%%%%%%%%%%%%%%%%%%%%%%%%%%%%%%%%

%%%%%%%%%%%%%%%%% BODY OF PAPER %%%%%%%%%%%%%%%%%%

\section{Introduction}

White dwarfs in close binaries (orbital periods less than a few days) with main sequence or substellar companions are the result of common envelope evolution \citep{Paczynski76,Webbink84}, in which the progenitor of the white dwarf expanded to fill its Roche lobe when it was a giant star and engulfed its companion. If the companion star was able to unbind the envelope of the giant before the system merged, then a post-common envelope binary emerges. These systems will continue to evolve to shorter periods as a result of angular momentum loss from magnetic braking \citep{Verbunt81,Rappaport83} and/or gravitational wave radiation \citep{Kraft62,Faulkner71}, eventually reaching the point where the companion star will fill its Roche lobe and transfer mass onto the white dwarf, forming a cataclysmic variable (CV).

Once formed, CVs continue to evolve to shorter periods due to angular momentum loss, with the donor star slowly stripped down to lower and lower masses. Eventually the donor will be reduced to a substellar mass object and the resulting structural change of the donor causes the binary to start evolving towards longer periods as a ``period bounce'' CV \citep{Knigge11,Belloni23}. Population synthesis calculations predict that the majority (40-60 per cent, \citealt{Goliasch15} or even as high as 75 per cent, \citealt{Belloni18}) of CVs should have already reached this evolutionary stage. However, observationally only around 7-14 per cent of CVs are period bouncers \citep{Pala20} and dedicated surveys for these objects appear to confirm that they are much rarer than predicted \citep{Inight23}, calling into question our understanding of the evolution of CVs.

\citet{Schreiber23} recently suggested a possible solution to this problem. Based on earlier work from \citet{Isern17} and \citet{Schreiber21}, they showed that if a magnetic field were to emerge from the white dwarf once the system had become a period bouncer, then the resulting transfer of angular momentum from the spin of the white dwarf back into the orbit, would cause the orbital separation to increase and result in the system detaching and no longer appearing as a CV. There is clear evidence that magnetism appears long after the formation of most white dwarfs \citep{Bagnulo21,Bagnulo22}, whether this is the result of a crystallization-powered dynamo or the emergence of a previously buried fossil field remains unclear \citep{Castro24,Camisassa24}. In either case the result is the same (i.e. the system will detach), if the field emerges after the white dwarf has already been spun up via mass transfer in a non-magnetic CV.

\citet{Schreiber23} showed that even a relatively weak magnetic field of just one MG could potentially detach systems for more than a Gyr, resulting in a 60-80 per cent decrease in the population of period bounce CVs, much more consistent with observations. However, as yet there is no direct evidence that this scenario is correct. A key prediction of this model is that there should be a large population of detached magnetic white dwarf plus brown dwarf binaries and the identification of such a population would strongly support this model. However, such a population is challenging to identify, since these white dwarfs will be cool and faint, while the brown dwarf masses could be as low as a few 10s of Jupiter masses \citep{Schreiber23,Schreiber25}, making them extremely faint and easy to hide, even next to a cool white dwarf. One way to identify such systems is through their eclipses, if they happen to have highly inclined orbits. When the larger, but much fainter, brown dwarf transits in front of the white dwarf it can result in deep eclipses that are easy to identify in multi-epoch photometric surveys \citep[e.g.][]{Kosakowski22}. In this paper we present the discovery of such a system and discuss whether it could previously have been a CV that detached as a result of the emergence of the magnetic field of the white dwarf.

\section{Discovery}

\begin{figure*}
  \begin{center}
    \includegraphics[width=\textwidth]{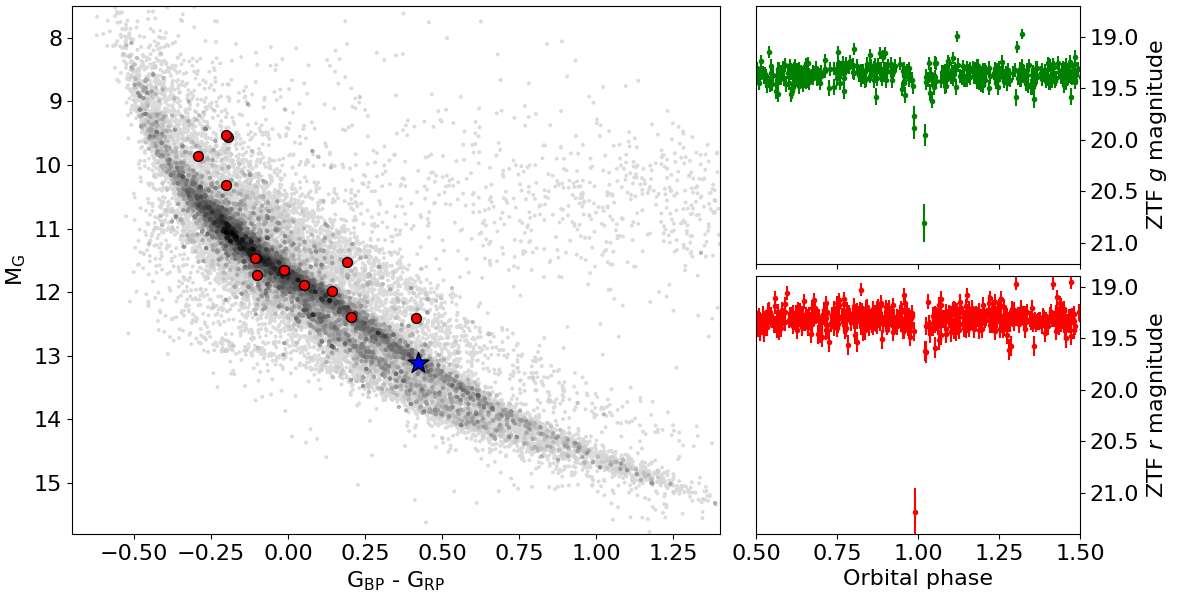}
    \caption{{\it Left:} Gaia DR3 colour-magnitude diagram of the white dwarf cooling track. Red points show confirmed close, detached white dwarf plus brown dwarf binaries \citep{Farihi04,Farihi17,Maxted06,Casewell12,Casewell18,Casewell20,Beuermann13,Parsons17b,Parsons25,Roestel21}, ZTF\,J0218+0711 is shown as a blue star, demonstrating that the white dwarf is considerably cooler than in other known systems and shows no clear optical excess from the companion to the white dwarf. {\it Right:} phase-folded ZTF light curves of ZTF\,J0218+0711 in the $g$ band (top) and $r$ band (folded on a period of 1.7 hours). The light curve is consistent with no variability outside of the deep eclipse of the white dwarf.}
  \label{fig:gaia_ztf}
  \end{center}
\end{figure*}

ZTF\,J021804.16+071152.93 (GALEX\,J021804.1+071152 in SIMBAD, hereafter ZTF\,J0218+0711) was first identified as a white dwarf from its Sloan Digital Sky Survey (SDSS) spectrum and listed as a DC white dwarf \citep{Kleinman13}. While the SDSS spectrum appears mostly featureless, the fitted temperature was $\sim$10,000\,K, significantly hotter than expected for a hydrogen atmosphere DC white dwarf. Similar ``hot'' DC white dwarfs have turned out to host strong magnetic fields, that distort and blur out their absorption lines, making them appear as DC white dwarfs in a noisy, low resolution spectrum \citep[e.g.][]{Parsons13}, offering the first evidence that ZTF\,J0218+0711 may be a magnetic white dwarf.

Gaia DR3 gives a parallax for ZTF\,J0218+0711 of $5.9\pm0.3$\,mas \citep{Gaia23}, implying a distance of around 172\,pc \citep{BailerJones21}. When placed on a colour-magnitude diagram, ZTF\,J0218+0711 sits towards the cooler end of the white dwarf cooling curve (see the left-hand panel of Figure~\ref{fig:gaia_ztf}), confirming that its temperature is likely lower than the fit to the SDSS spectrum implies. The fact that ZTF\,J0218+0711 appears to sit exactly on the cooling track would indicate that this is likely an isolated white dwarf. However, inspection of its Zwicky Transient Facility (ZTF) light curve \citep{Bellm19,Masci19} reveals a deep eclipse of the white dwarf every 1.7 hours (see the right-hand panel of Figure~\ref{fig:gaia_ztf}). These eclipses were found through a dedicated search for eclipsing white dwarfs in ZTF using a box least squares analysis, the full details of which will be presented in a forthcoming paper (van Roestel et~al. in preparation).

The boxy-shaped eclipse seen in the ZTF light curves implies that the white dwarf is eclipsed by a much larger object, ruling out a double white dwarf system. The fact that this cool and faint white dwarf shows no clear optical red excess means that the companion to the white dwarf must be extremely faint and almost certainly substellar, prompting the follow up observations presented in this paper.

\section{Observations and their reduction}

\begin{table*}
 \centering
  \caption{Journal of observations.}
  \label{tab:obslog}
  \begin{tabular}{@{}lcccccc@{}}
  \hline
  Telescope/ & Date & Filters & No. of    & Exposure  & Transparency & seeing \\
  Instrument &      &         & exposures & times (s) &             & (arcsec) \\
  \hline
  VLT/X-shooter & 2024-09-02 & UVB,VIS,NIR & 10,9,46   & 588,600,120 & Clear & 0.5--0.7 \\
  GTC/HiPERCAM  & 2023-09-17 & $u_s$,$g_s$,$r_s$,$i_s$,$z_s$ & 357,1072,1072,536,536 & 7.5,2.5,2.5,5.0,5.0 & Clear & 0.7--9.0 \\
  VLT/FORS2 & 2024-10-30 & - & 14 & 450 & Clear & 1.0--1.2 \\
  \hline
\end{tabular}
\end{table*}

In this section we summarise our observations and their reduction. A full list of our observations is given in Table~\ref{tab:obslog}.

\subsection{X-shooter spectroscopy}

ZTF\,J0218+0711 was observed with the echelle spectrograph X-shooter \citep{Vernet11} mounted at the Cassegrain focus of the VLT-UT3 at Paranal, Chile, on the night of the 2024 September 2. X-shooter provides medium resolution spectroscopic data from 300-550\,nm in the UVB arm, 550-1000\,nm in the VIS arm and 1-2.4\,microns in the NIR arm. The observations of ZTF\,J0218+0711 were performed in STARE mode (the telescope was not nodded between exposures). This approach, combined with the faintness of the target at infrared wavelengths, means that the NIR arm data have extremely low signal-to-noise ratios and we do not include any of these data in our subsequent analysis. 

We used slit widths of 1.0 arcsec in the UVB arm and 0.9 arcsec in the VIS arm, binned the detector by a factor of two in the spatial direction in both arms and by a factor of two in the dispersion direction of the UVB arm. This results in a resolution of R$\simeq$5000 in the UVB and R$\simeq$8000 in the VIS arm. 

All of the data were reduced using the standard X-shooter pipeline release (version 3.6.3) within {\sc esoreflex} \citep{Freudling13}. The standard X-shooter pipeline can reach a velocity accuracy of around 8\,{\kms} in the VIS arm. We improved this accuracy by fitting a number of telluric absorption features and corrected for small (typically $\sim$1\,{\kms}) systemic velocity offsets in the data via the method described in \citet{Parsons17}. This allowed us to reach an accuracy of a few {\kms} around the H$\alpha$ line. All spectra were then placed on a barycentric wavelength scale.

\subsection{HiPERCAM photometry}

High-speed multi-band light curves of ZTF\,J0218+0711 were obtained with HiPERCAM \citep{Dhillon21} mounted on the 10.4-m Gran Telescopio Canarias (GTC) on La Palma, Spain on 2023 September 17. HiPERCAM allows simultaneous observations in the super-SDSS $u_s$,$g_s$,$r_s$,$i_s$ and $z_s$ bands with negligible (8\,ms) dead-time between exposures. We used exposure times of 2.5 seconds in the $g_s$ and $r_s$ bands, 5 seconds in the $i_s$ and $z_s$ bands and 7.5 seconds in the $u_s$ band.

The HiPERCAM data were reduced using the HiPERCAM pipeline software\footnote{\url{https://github.com/HiPERCAM/hipercam}}. After bias subtraction and flat fielding (using twilight flats) the source flux was determined with aperture photometry using a variable aperture scaled according to the full width at half maximum (FWHM). The HiPERCAM $z_s$ band data were also defringed using a reference fringe map. Transmission variations were accounted for by determining the flux relative to a non-variable comparison star in the field of view. Unfortunately ZTF\,J0218+0711 was considerably brighter than all other stars in the field of view in the $u_s$ band, meaning that no suitable comparison star was available in the $u_s$ band. We combined the flux of several nearby stars to use as a comparison source, but the final light curve is still dominated by the noise in these comparison stars, rather than the source itself. Fortunately, given the lower time resolution of the $u_s$ band data, these data are not essential for determining the system parameters (the higher time resolution and signal-to-noise ratio of the redder bands help constrain the stellar and binary parameters far better than the $u_s$ band data). The $u_s$ band data only help constrain the temperature of the white dwarf, in combination with the other bands (see section~\ref{sec:lcfit}).

The data were flux calibrated using observations of the standard star G\,184$-$17, based on the reference magnitudes for the super-SDSS filters from \citet{Brown22}. Finally, all times were converted to a TDB (Barycentric Dynamical Time) timescale corrected to the barycentre of the solar system. Using modified Julian dates format, this results in timing measurements of the form BMJD(TDB).

\subsection{FORS2 spectropolarimetry}

\begin{figure*}
    \begin{center}
        \includegraphics[width=0.9\textwidth]{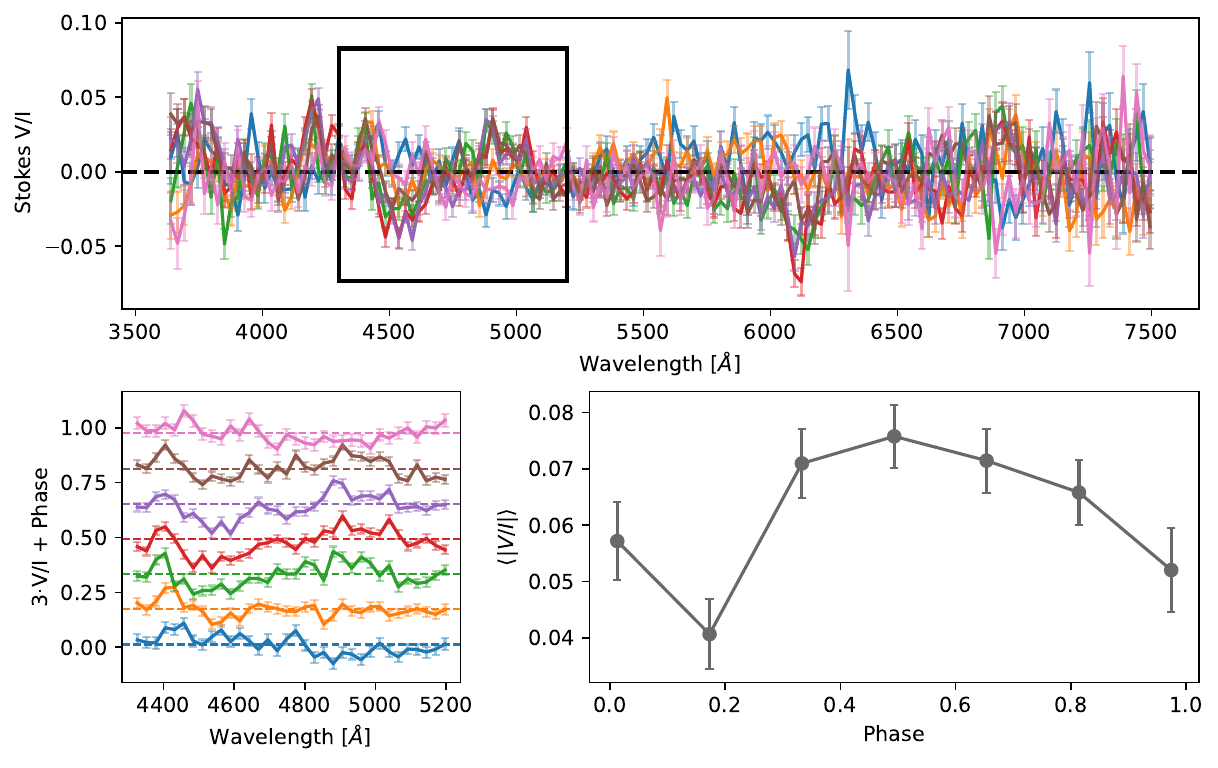}
    \end{center}
    \caption{\label{fig:specpol} Circular polarisation (V/I) spectra of 7 observations of ZTF\,J0218+0711 showing variability, which is caused by the rotation of the white dwarf. The top panel shows all spectra overlaid over the full wavelength range, with the black box marking the area of clearest variability. The bottom-left panel shows the polarised spectra corresponding to the box in the top panel in phase with the orbital ephemeris. The bottom-right panel shows the mean of the absolute polarisation values in phase with the orbital ephemeris. The data appears periodic on a timescale consistent with the orbital period, however, a clear period cannot be determined given the limited baseline of the observations.}
\end{figure*}

Spectropolarimetric observations were obtained with the FORS2 instrument \citep{Appenzeller98} at the ESO VLT, using the 300V grism and a 1.2 arcsec slit, corresponding to a resolving power of $\sim$ 360. We acquired a sequence of 14 exposures of 450 s each, with the retarder waveplate set at position angles of $-45$, $+45$, $+45$, $-45$, etc. on 2024 October 30 between UT 03:11:43 and UT 05:05:56. The observations were carried out using the blue-optimised 2k$\times$4k E2V CCD.

The data were recombined in pairs of frames obtained at waveplate position angles separated by $90^\circ$, following the standard beam-swapping technique \citep[e.g.][]{Bagnulo09}. This procedure yielded a set of seven Stokes $I$ and Stokes $V/I$ spectra, with Stokes $I$ not flux-calibrated. The data reduction was performed using the FORS pipeline \citep{Izzo10} to produce wavelength-calibrated and rectified 2D frames, followed by beam extraction using IRAF routines.

\section{Results}

\subsection{Spectropolarimetry analysis}

The FORS2 observations of ZTF\,J0218+0711 show clear variability in the circular polarisation of several spectral features, as can be seen in the top panel of Figure~\ref{fig:specpol}. The detection of circular polarisation from ZTF\,J0218+0711 confirms the magnetic nature of the white dwarf. The variability is shown in more detail in the bottom two panels for the 4300--5300\,{\AA} region, where the left panel shows the polarised spectra in phase with the orbital ephemeris of the system, and the average of the absolute polarisation values on the right. 

The periodicity of the observed variability was analysed through simultaneous sinusoidal fitting of all wavelength bins $i$ in a given wavelength range. The best fit is determined by minimising $\chi^2$, which is expressed as
\begin{equation}
    \chi^2 = \sum_{i=1}^N\,\frac{\left((V/I)_i - (A_i\sin(\frac{2\pi}{P}t + \phi_i) + \mu_i)\right)^2}{\sigma_{V/I,i}^2},
\end{equation}
where $(V/I)_i$ and $\sigma_{V/I,i}^2$ are the observed polarisation value and uncertainty in wavelength bin $i$, $N$ the total number of bins, and $P$ the rotational period of the white dwarf. Each bin is modelled as a sine wave with period $P$ and individual values for amplitude $A_i$, mean $\mu_i$, and phase $\phi_i$. The best fit period from analysing either the full wavelength range or specific ranges, such as 4300--5300\,{\AA}, is on the order of several hours with a lower bound of 90 minutes and no clear upper bound. While the variability visually appears to be periodic on a timescale consistent with the orbital period (implying that the spin of the white dwarf is locked to the orbit), the rotational periodicity cannot be determined more clearly due to the short baseline of 100 minutes, which only covers a single orbital period and so we cannot rule out a longer rotational period for the white dwarf.

\subsection{Spectroscopic analysis}

\begin{figure}
  \begin{center}
    \includegraphics[width=\columnwidth]{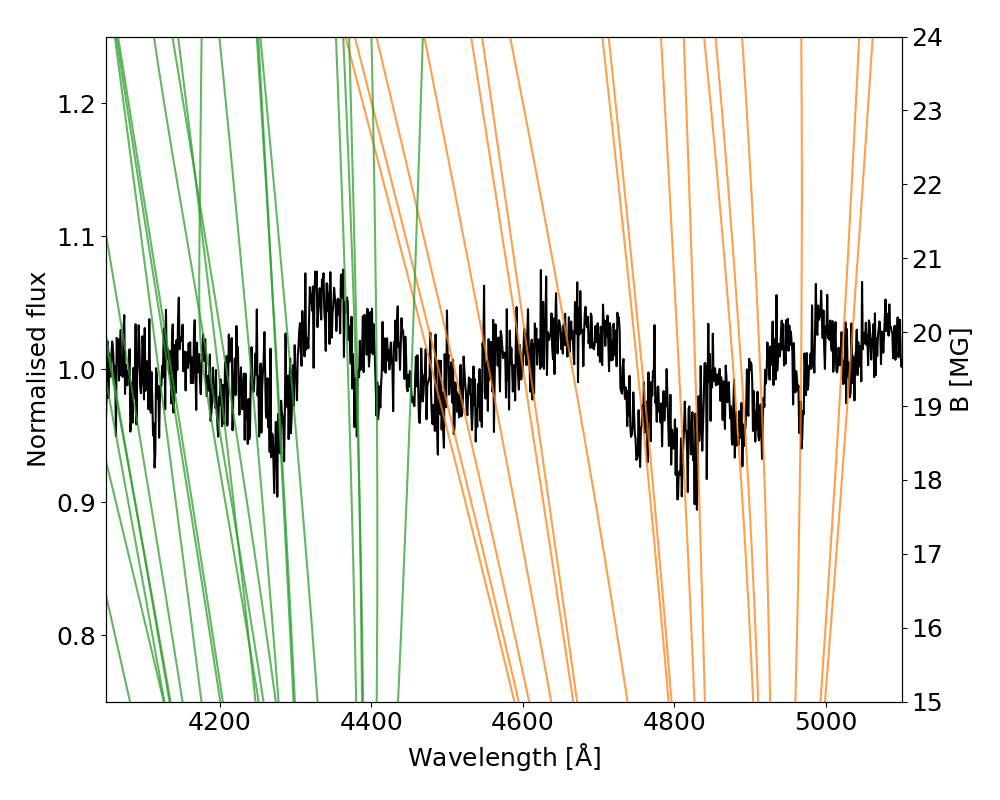}
    \caption{Normalised, averaged X-shooter spectrum of ZTF\,J0218+0711 binned by a factor of 5 (black line). Overplotted are the Zeeman split components as a function of the magnetic field strength for the H$\beta$ line (orange) and H$\gamma$ (green). The dips in the spectrum agree well with a field strength of around 19\,MG.}
  \label{fig:zeeman}
  \end{center}
\end{figure}

The X-shooter spectra show a mostly featureless, blue continuum from the white dwarf. However, when the individual spectra are combined, the higher signal-to-noise ratio reveals a number of absorption features. Given that the white dwarf completely dominates the optical light, these lines must originate from the white dwarf, but do not appear consistent with a typical DA white dwarf spectrum. Figure~\ref{fig:zeeman} shows a zoom in on the region around the H$\gamma$ and H$\beta$ lines highlighting these multiple absorption features, which are consistent with the Zeeman split components of the Hydrogen Balmer lines in a roughly 19\,MG magnetic field. This further confirms the magnetic nature of the white dwarf and can be used to get a good estimate of the surface field strength. While these features are also visible in the individual spectra the signal-to-noise ratio is far too low to detect the $\sim$20\,{\kms} radial velocity variations from the white dwarf. Moreover, the changing orientation of the magnetic field as the white dwarf rotates will lead to far larger shifts in the positions of these lines than the actual velocity of the white dwarf, meaning that we are unable to constrain the white dwarf's radial velocity from the X-shooter data.

\begin{figure*}
  \begin{center}
    \includegraphics[width=\textwidth]{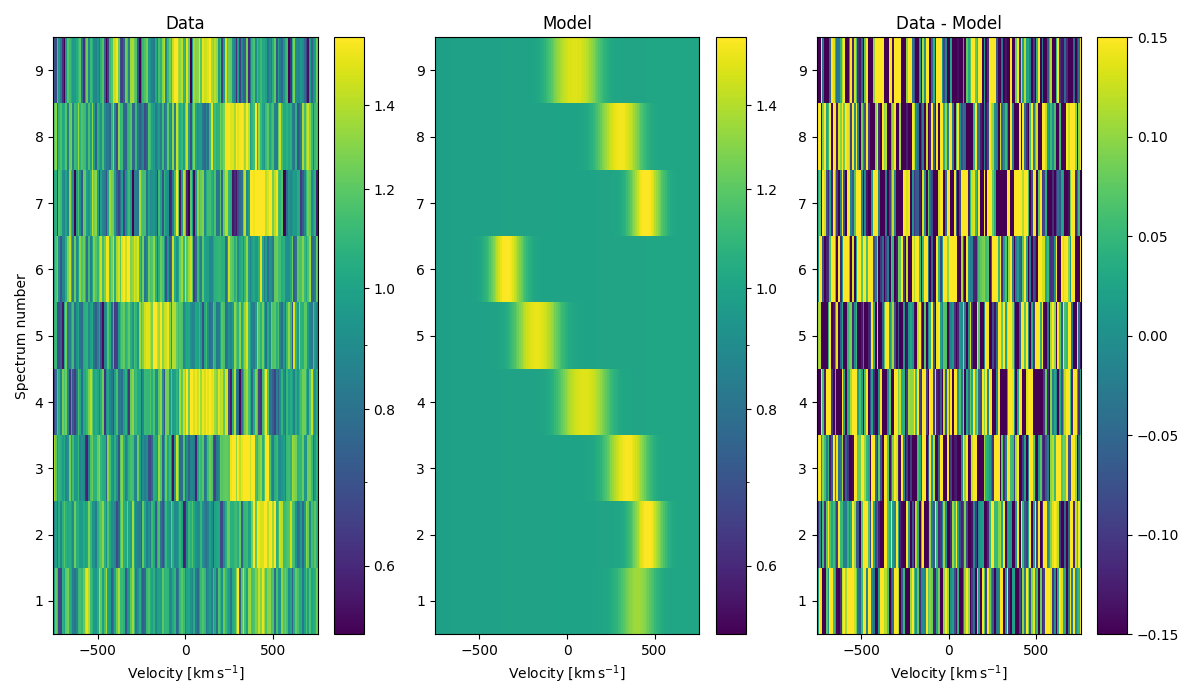}
    \caption{Trailed spectra of the H$\alpha$ emission line originating from the companion of the white dwarf in ZTF\,J0218+0711. {\it Left:} original data, where there is a time jump between spectrum 6 and 7, {\it middle:} our best fit model to the data, {\it right:} the residuals of the fit. Colours are normalised flux density, as shown by the colourbars.}
  \label{fig:trail}
  \end{center}
\end{figure*}

In addition to these absorption lines from the white dwarf, the X-shooter data also revealed a narrow H$\alpha$ emission component that moves with high velocity (see the left-hand panel of Figure~\ref{fig:trail}). Based on the high velocity and orbital phase information it is clear that this emission originates from the companion star to the white dwarf. H$\alpha$ emission is commonly seen in white dwarf plus main sequence binaries \citep{Rebassa07} and is usually the result of chromospheric emission from the main sequence star, which generally tracks the centre of mass of the star \citep[e.g.][]{Parsons17}. White dwarfs with substellar companions can also show emission from the companion, but in these cases the emission is the result of irradiation of the substellar object by the much hotter white dwarf \citep[e.g.][]{Parsons17b}. Since only one side of the substellar object is irradiated, the emission is localised on the inner hemisphere and hence tracks the centre of light of this emission region, rather than the centre of mass of the object itself (i.e. it is effectively a lower limit on the true radial velocity of the substellar object). In the case of ZTF\,J0218+0711 this emission is somewhat unexpected, given the low temperature of the white dwarf. In irradiated systems the companion star absorbs the ultraviolet light of the white dwarf, which is then re-emitted as optical and infrared light. The white dwarf in ZTF\,J0218+0711 has a temperature of only 8500\,K (see Section~\ref{sec:lcfit}) and hence emits very little ultraviolet light. All other white dwarfs with substellar companions that show emission lines have white dwarfs hotter than $\sim$13,000\,K\footnote{NLTT\,5306 has a 7750\,K white dwarf and shows H$\alpha$ emission, but this originates from the white dwarf and not the substellar object \citep{Steele13}}. The fact that the white dwarf in ZTF\,J0218+0711 also hosts a strong magnetic field may help explain the origin of this emission. With an orbital separation of only 0.65\,{\RSUN} (see Section~\ref{sec:lcfit}) the strength of the white dwarf's magnetic field is likely to be 10s if not 100s of kG at the surface of the companion. This is likely to be far larger than the intrinsic magnetic field of the companion \citep{Reiners10} and therefore may well induce the observed emission, meaning that the origin of this emission could be quite different to the standard irradiation model.

In order to measure the radial velocity semi-amplitude of this emission line we fitted the data following the method outlined in \citet{Parsons17b}, in which the H$\alpha$ line is fitted with a combination of a first order polynomial (to represent the underlying continuum) and a Gaussian emission component. All of the spectra are fitted simultaneously, with the position of the emission component changing according to $\gamma_\mathrm{BD}$ + $K_\mathrm{emis} \sin{(2 \pi \phi)}$, where $\phi$ is the orbital phase ($\phi=0$ is the centre of the eclipse of the white dwarf). The width of the Gaussian component remains the same in all spectra, while its height is modulated as $1 - \cos({2 \pi \phi})$, to simulate the increased strength of this component as the inner hemisphere becomes more visible (i.e. the line is strongest at $\phi=0.5$ and weakest at $\phi=0$).

We determined the orbital phase of each spectrum using the orbital period ($P_\mathrm{orb}$) measured from the ZTF light curve and the time of mid-eclipse of the white dwarf ($T_0$) from our HiPERCAM observations (see Section~\ref{sec:lcfit}). The orbital phase is then given by:
\begin{equation}
\phi = (T_\mathrm{obs}-T_0)/P_\mathrm{orb},
\end{equation}
where $T_\mathrm{obs}$ is the mid-exposure time of the spectrum. The emission component moves sufficiently fast that its motion is smeared by the finite exposure times in the spectra taken around $\phi=0.5$ (where its velocity is changing fastest). To account for this smearing we divide each spectrum into 5 equal sub-exposures, covering the full phase range of each spectrum, and calculate a model for each sub-exposure, then average these together to create the final model for that spectrum. The simultaneous fit to all spectra was performed using the Markov chain Monte Carlo (MCMC) method as implemented in the {\sc emcee} Python package \citep{Foreman13}. We used 100 walkers, with a burn-in period of 1500 and 10 000 production steps, with convergence checked using the Gelman-Rubin statistic \citep{Gelman92}.

The best fit model of the H$\alpha$ emission line is shown in the centre panel of Figure~\ref{fig:trail} and the residuals to the fit are in the right-hand panel of the same figure. We find $\gamma_\mathrm{BD}=43.8 \pm 7.1$\,{\kms} and $K_\mathrm{emis} = 409.8 \pm 5.7$\,{\kms}.

As previously noted, the radial velocity of the emission line is unlikely to track the true centre of mass of the companion star. The relationship between the radial velocity semi-amplitude of the emission line ($K_\mathrm{emis}$) and the centre of mass radial velocity semi-amplitude ($K_\mathrm{BD}$) is given by \citep{Wade88}:
\begin{equation}
K_\mathrm{BD} = \frac{K_\mathrm{emis}}{1 - f(1+q) R_\mathrm{BD}/a}, \label{eqn:kcorr}
\end{equation}
where $q=M_\mathrm{BD}/M_\mathrm{WD}$ is the mass ratio, $R_\mathrm{BD}/a$ is the radius of the companion star scaled to the semi-major axis and $f$ is is a constant between 0 and 1 which depends upon the location of the centre of light on the irradiated hemisphere \citep{Parsons12}. In the most extreme case ($f=1$) the emission is assumed to entirely originate at the point on the companion's surface closest to the white dwarf (the substellar point), while $f=0$ assumes that the emission is uniform over the surface of the star and hence does track the centre of mass of the star. Therefore, we can use this emission feature to help constrain the stellar and binary parameters, in combination with information from the light curve.

\subsection{Light curve analysis} \label{sec:lcfit}

\begin{figure}
  \begin{center}
    \includegraphics[width=\columnwidth]{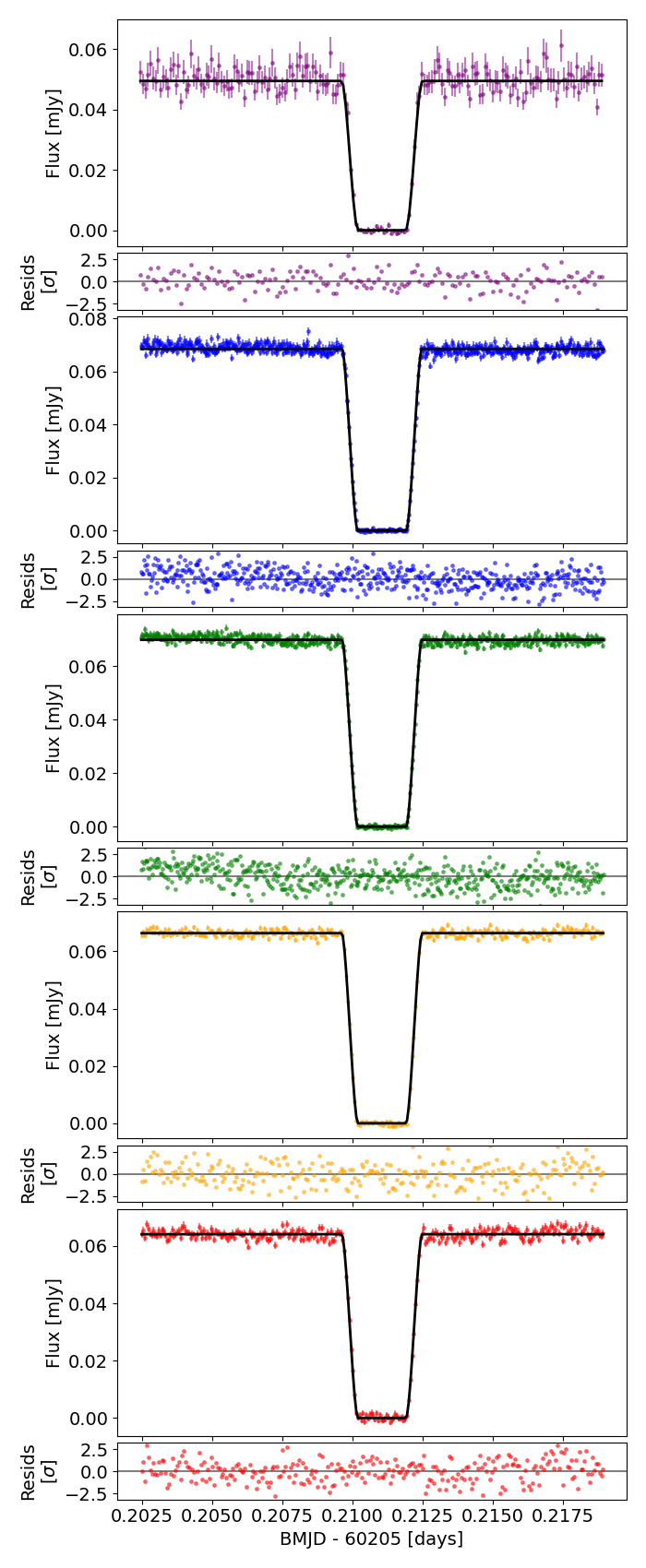}
    \caption{GTC+HiPERCAM eclipse light curves of ZTF\,J0218+0711 (top to bottom: $u_s$, $g_s$, $r_s$, $i_s$ and $z_s$ bands) along with the best fit models (black lines). Residuals to the fits are shown beneath each panel (in standard deviations).}
  \label{fig:lc_fit}
  \end{center}
\end{figure}

The HiPERCAM light curves only cover the orbital phases around the eclipse of the white dwarf, but show a very typical shape for a detached white dwarf plus substellar object system (see Figure~\ref{fig:lc_fit}). The eclipse is extremely deep in all bands, indeed we were unable to detect any flux from the companion during the eclipse in any of the bands. We determined upper limits on the brightness of the companion of $i_s > 24.0$ and $z_s > 23.7$, which correspond to absolute magnitudes of $M_i > 17.8$ and $M_z > 17.5$ (based on the Gaia DR3 parallax), implying a spectral type later than L4 \citep{Kiman19}. 

Interestingly, despite the 19\,MG magnetic field of the white dwarf and the short period, there are no clear signs of magnetism in the light curves. Typically, even low levels of wind accretion onto the white dwarf are sufficient to drive strong cyclotron emission that can significantly distort the shape of the eclipse in white dwarf plus main sequence star systems \citep[e.g.][]{Brown23}. Strong cyclotron emission is seen in magnetic white dwarf plus brown dwarf binaries as well \citep{Schmidt05,Schmidt07,Breedt12,Parsons21}, although it is unclear whether these systems are Roche lobe filling. In the case of ZTF\,J0218+0711 the companion is definitely not Roche lobe filling. Moreover, the field strength of only 19\,MG means that the cyclotron fundamental will be far in the infrared (around 59 microns) and we would require emission at the 6th harmonic or higher to get optical cyclotron emission, which is unlikely at the expected very low wind accretion rate in this system \citep{Woelk92}, if there is any accretion at all. It is therefore unsurprising that we do not see any strong cyclotron emission in any of our data. This also means that we could model the eclipse light curves assuming that all the light comes from the photosphere of the white dwarf, significantly simplifying the model.

The shape of the eclipse is related to the radii of the two stars and the orbital inclination. Moreover, the out of eclipse flux in each of the bands is related to the temperature and surface gravity of the white dwarf, plus the distance and reddening of the system. However, there is not enough information in the light curves alone to uniquely determine the values of all these parameters simultaneously, because of well known degeneracies between them \citep{Parsons17}. While the additional information from the H$\alpha$ emission line from the companion somewhat helps, because we do not have a direct measurement of the centre of mass radial velocity for either star, we cannot break these degeneracies. We therefore needed to rely on some theoretical relationships, in addition to the light curves, to constrain the stellar and binary parameters.

We used a similar approach to \citet{Brown22} to determine the stellar and binary parameters. Model eclipse light curves are generated using the {\sc lcurve} code \citep{Copperwheat10}. For a detached binary the main input parameters for {\sc lcurve} are the mass ratio ($q=M_\mathrm{BD}/M_\mathrm{WD}$), the orbital inclination ($i$), the radii of the two stars scaled by the orbital separation ($R_\mathrm{WD}/a$ and $R_\mathrm{BD}/a$), the temperatures of the two stars ($T_\mathrm{WD}$ and $T_\mathrm{BD}$), the orbital period ($P_\mathrm{orb}$), time of mid-eclipse ($T_0$) and limb darkening parameters for both stars. 

To break the degeneracy between the orbital inclination and scaled radii of the two stars we forced the white dwarf to follow the theoretical mass-radius relationship of \citet{Bedard20} for a C/O core white dwarf. In addition to this, we calculated the white dwarf flux in each of the HiPERCAM bands following the approach of \citet{Brown22}, which uses the tables from \citet{Claret20} to determine the emergent flux from the white dwarf, which is then scaled based on distance and extinction. These model fluxes are then compared to the observed out of eclipse flux in the HiPERCAM light curves. This approach allowed us to constrain the temperature of the white dwarf from the light curves (albeit based on non-magnetic DA white dwarf models) and gave an additional constraint on the surface gravity of the white dwarf. For the temperature and magnetic field strength of this white dwarf the difference between magnetic and non-magnetic models are a few hundredths of a magnitude \citep{Hardy23}, comparable to the uncertainty on the measurements and therefore we expect this to have a negligible effect on the fit.

Our full fit therefore combined {\sc lcurve} model fits to the eclipse profile and white dwarf model fits to the out of eclipse fluxes (effectively a spectral energy distribution fit). For a more complete description of this method see \citet{Brown22}. The final input parameters for the fit were $M_\mathrm{WD}$, $M_\mathrm{BD}$, $R_\mathrm{BD}$, $T_\mathrm{WD}$, $i$, $T_0$, the parallax ($\varpi$, from which the distance was calculated) and reddening ($E(B-V)$, from which the extinction in each band was calculated). From these input parameters, plus Kepler's third law and the white dwarf mass-radius relationship from \citet{Bedard20} we calculated all of the required {\sc lcurve} inputs. Since the companion is not detected in any of the light curves the data are insensitive to the values of $T_\mathrm{BD}$ and the limb darkening for the companion, so these were fixed at 1000\,K and a linear limb darkening coefficient of 0.5 (these choices have no impact on the final results). Since our HiPERCAM observations only covered one eclipse we fixed $P_\mathrm{orb}$ to the value found from the ZTF light curve. We determined the limb darkening coefficients for the white dwarf by interpolating the tables from \citet{Claret20}, using the four-term law.

\begin{figure}
  \begin{center}
    \includegraphics[width=\columnwidth]{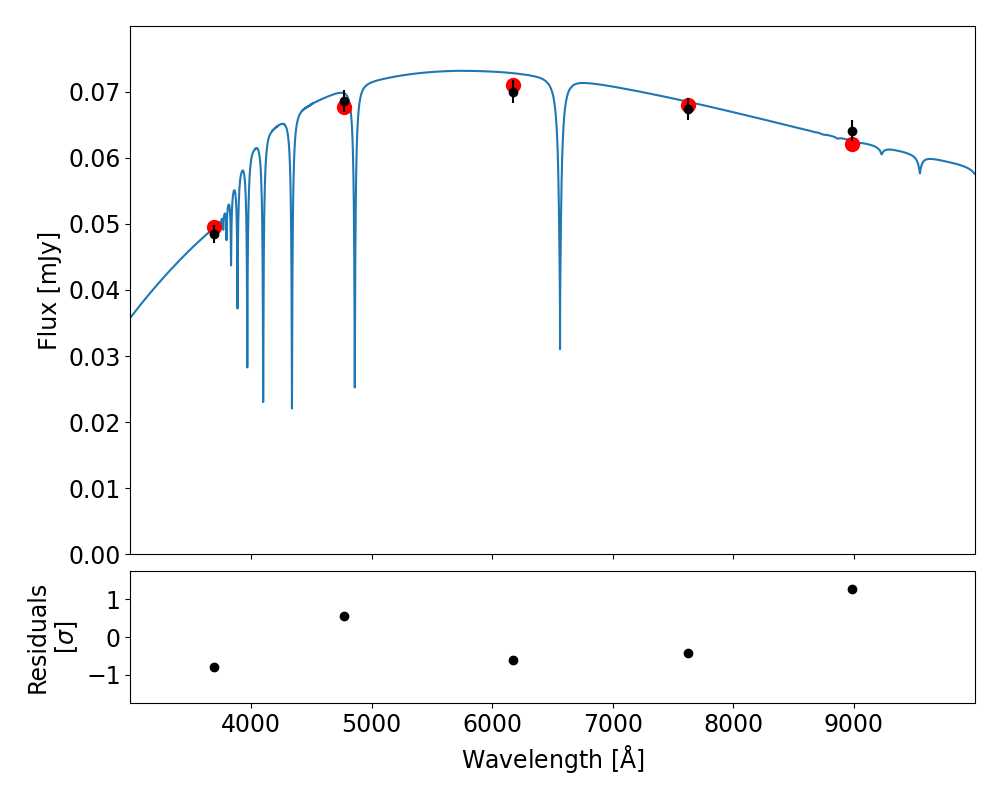}
    \caption{SED fit of the white dwarf in ZTF\,J0218+0711, based on the fit to the GTC+HiPERCAM eclipse light curves. The black points are effectively the out of eclipse fluxes in the HiPERCAM $u_s$, $g_s$, $r_s$, $i_s$ and $z_s$ bands. A (non-magnetic) model white dwarf spectrum is shown, based on the best fit parameters. The red points show the model fluxes in the HiPERCAM bands. Residuals to the fit are shown in the lower panel (in standard deviations).}
  \label{fig:sed_fit}
  \end{center}
\end{figure}

For the fit we placed a Gaussian prior on the parallax, based on the Gaia DR3 value \citep{Gaia23} and a uniform prior on the reddening (between 0 and 0.01) based on the STILISM reddening map \citep{Capitanio17}. Additionally, for each model generated during the fit we also calculated the value of $K_\mathrm{emis}$ via Equation~\ref{eqn:kcorr} and Kepler's third law:
\begin{equation}
\frac{P_\mathrm{orb} K_\mathrm{BD}^3}{2 \pi G} = \frac{M_\mathrm{WD} \sin^3 i}{(1+q)^2},
\label{eqn:k3}
\end{equation}
and placed a Gaussian prior on this value based on the measured value and uncertainty from the X-shooter spectra. This ensured that our light curve fit was also consistent with the spectroscopic results and also helped to constrain the masses, inclination and the size of the companion. However, in order to calculate $K_\mathrm{emis}$ we needed to assume a value of $f$ in Equation~\ref{eqn:kcorr}. Typically, for irradiation-driven H$\alpha$ emission in white dwarf binaries a value of $f=0.5$ is assumed \citep[e.g.][]{Parsons17}. However, since it is unclear exactly how the H$\alpha$ emission is generated in this system it is not obvious that this is the correct choice. Nevertheless, in the absence of any additional constraints we decided to fix $f=0.5$ for our fit (there is not enough information in the data to allow this value to be free). We also repeated our fit with the two extreme values of $f=0$ and $f=1$, to see what impact this had on the final parameters.

\begin{figure*}
  \begin{center}
    \includegraphics[width=\textwidth]{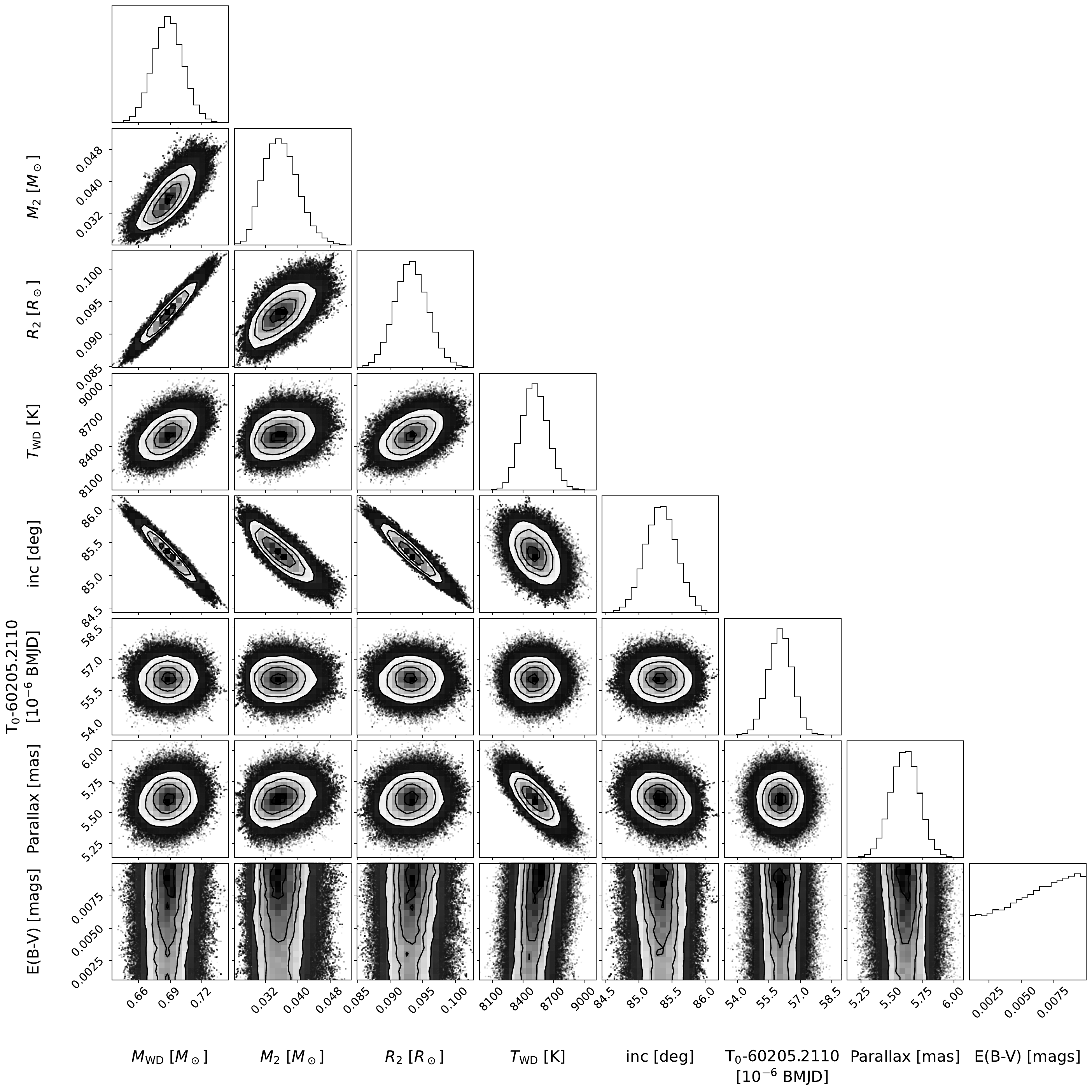}
    \caption{Posterior probability distributions for model parameters obtained through fitting the GTC+HiPERCAM light curves of ZTF\,J0218+0711.}
  \label{fig:1828_lcurve_corner}
  \end{center}
\end{figure*} 

The fit was performed using the MCMC method, where the goodness of fit was determined from a combination of the fit to the five HiPERCAM lightcurves and the out of eclipse flux measurements (see section 2.6 of \citealt{Brown22} for a more detailed explanation). We used 100 walkers and 20 000 steps, with the first 2000 discarded as burn-in. The best fit light curve models are shown in Figure~\ref{fig:lc_fit}, the SED fit is shown in Figure~\ref{fig:sed_fit} and the posterior probability distributions of the parameters are shown in Figure~\ref{fig:1828_lcurve_corner}. The best fit values and their uncertainties are also listed in Table~\ref{tab:params}, based on the median and 16th and 84th percentiles of the probability distributions.

The fit revealed that the white dwarf has a fairly typical mass for an isolated white dwarf ($0.69 \pm 0.01$\,\MSUN), but somewhat above average for white dwarfs in post-common envelope binaries \citep{Rebassa11}. With an effective temperature of only 8500\,K this is the coolest white dwarf with an eclipsing substellar companion, with the exception of the transiting planet around WD\,1856+534 \citep{Vanderburg20}. The fit also confirms that the companion is indeed substellar, with a mass of only $0.036 \pm 0.005$\,{\MSUN} ($37 \pm 5$\,{\MJUP}). This is a little lower than the typical mass of around 50\,{\MJUP} seen in many white dwarf plus brown dwarf binaries, but systems are known with lower mass companions \citep[e.g.][]{Parsons25}. The brown dwarf is also relatively far from filling its Roche lobe, with a fill factor of only 0.78.

We re-ran the fit twice, changing the $f$ factor in Equation~\ref{eqn:kcorr} to 0 and 1 to determine the effect that this has on our final results. Many parameters are unaffected by the choice of this value or vary by an amount smaller than the uncertainties quoted in Table~\ref{tab:params}. Decreasing the value to $f=0$ (i.e. assuming that the H$\alpha$ emission line traces the centre of mass of the brown dwarf) had the largest effect, decreasing the white dwarf mass to 0.630\,{\MSUN}, while the mass of the brown dwarf increases to 0.045\,{\MSUN} (47\,{\MJUP}). Increasing the value to $f=1$ (i.e. assuming that the H$\alpha$ emission line only arises from the point on the brown dwarf closest to the white dwarf) had a marginal effect, with the mass of the white dwarf slightly increasing to 0.695\,{\MSUN} and the mass of the brown dwarf decreasing slightly to 0.030\,{\MSUN} (31\,{\MJUP}). The true masses therefore sit somewhere between these extremes, but in the following discussions we adopt the values from the $f=0.5$ fit shown in Table~\ref{tab:params}, although we note that our conclusions remain the same, regardless of what values are chosen for the masses of the two objects.

\section{Discussion}

\begin{table}
 \centering
  \caption{Stellar and binary parameters for ZTF\,J0218+0711. $K_\mathrm{emis}$ is the radial velocity semi-amplitude of the H$\alpha$ emission line from the brown dwarf, not the centre of mass velocity of the brown dwarf. The white dwarf is forced to follow the mass-radius relationship of \citet{Bedard20}. RLFF is the volume-average Roche lobe fill factor}
  \label{tab:params}
  \tabcolsep=0.5cm
  \begin{tabular}{@{}lcc@{}}
  \hline
  Parameter & Unit & Value \\
  \hline
  \multicolumn{3}{l}{From Gaia DR3 \citep{Gaia23}:}\\
  RA & hours              & 02:18:04.16 \\
  Dec & deg               & +07:11:52.93 \\
  $\varpi$ & mas          & $5.87 \pm 0.32$ \\
  $\mu_{\alpha} \cos \delta$ & mas/yr & $10.1 \pm 0.3$ \\
  $\mu_{\delta}$ & mas/yr & $-76.2 \pm 0.3$ \\
  $G$ & mag          & $19.266 \pm 0.005$ \\
  $G_\mathrm{BP}-G_\mathrm{RP}$ & mag & $0.421 \pm 0.074$ \\
  \multicolumn{3}{l}{From \citet{BailerJones21}:}\\
  $D$& pc                 & $172^{+7}_{-9}$ \\
  \multicolumn{3}{l}{From ZTF light curves:}\\
  $P_\mathrm{orb}$ & days & 0.070\,969\,585\,59(50) \\
  \multicolumn{3}{l}{From X-shooter spectra:}\\
  $K_\mathrm{emis}$ & \kms  & $409.8 \pm 5.7$ \\
  $\gamma_\mathrm{BD}$ & \kms & $43.8 \pm 7.1$ \\
  $B_\mathrm{WD}$ & MG    & $\simeq$19 \\
  \multicolumn{3}{l}{From HiPERCAM eclipse light curves:}\\
  $T_0$ & BMJD(TDB)       & 60205.211\,056\,04(57) \\
  $i$ & deg               & $85.3 \pm 0.3$ \\
  $M_\mathrm{WD}$ & \MSUN & $0.688 \pm 0.014$ \\
  $T_\mathrm{eff,WD}$ & K & $8510 \pm 130$ \\
  $M_\mathrm{BD}$ & \MSUN & $0.036 \pm 0.005$ \\
  $R_\mathrm{BD}$ & \RSUN & $0.093 \pm 0.003$ \\
  \multicolumn{3}{l}{Derived:}\\
  $a$ & \RSUN             & $0.648 \pm 0.005$ \\
  WD $\log(g)$ & cgs      & $8.141 \pm 0.021$ \\
  $R_\mathrm{WD}$ & \RSUN & $0.0117 \pm 0.0002$ \\
  RLFF            & -     & $0.78 \pm 0.04$ \\
  \hline
\end{tabular}
\end{table}

\subsection{A thick disk star?}

We analysed the kinematics of ZTF\,J0218+0711 to attempt to identify which galactic population it belongs to. The relatively large proper motion and systemic velocity ($\gamma_\mathrm{BD}$) makes it stand out as a potential member of an older galactic population, which can place useful constraints on the total age of the system. We calculated the space velocity of ZTF\,J0218+0711 relative to the Sun using the Gaia DR3 coordinates and proper motion components, the geometric distance from \citet{BailerJones21} and the systemic velocity measured from the X-shooter data. We used {\sc astropy} \citep{Astropy13,Astropy18,Astropy22} for these calculations. A right-handed Galactocentric frame is assumed, with a solar position of $x_\odot$ = (-8.12, 0.00, 0.02)\,kpc, and solar velocity of
$v_\odot$ = (12.9, 245.6, 7.8)\,{\kms} \citep{Reid04,Drimmel18,GRAVITY19}. We calculated space velocity components for ZTF\,J0218+0711 of $v_{\phi} = -6.1 \pm 5.7$\,{\kms}, $v_R = -39.4 \pm 6.7$\,{\kms} and $v_z = -65.3 \pm 7.3$\,{\kms}. While these values are not extremely unusual, the large $v_z$ component is noteworthy, as it is quite inconsistent with a thin disk star. Indeed, following the approach of \citet{Bensby03} we find that these space velocity components imply that ZTF\,J0218+0711 is a high probability member of the galactic thick disk.

To investigate this further, we integrated the galactic orbit of ZTF\,J0218+0711 using the {\sc gala} package \citep{Price17}, using a standard Milky Way potential model. We found that the system is on a relatively circular orbit around the Galaxy ($e=0.09$) but reaches a significant distance above the galactic plane ($Z_\mathrm{max}=1.2$\,kpc), much larger than the $\sim$0.25\,kpc scale height of the thin disk \citep{Imig25}. While a metallicity measurement from the brown dwarf component would help confirm this, we nevertheless conclude that ZTF\,J0218+0711 is highly likely to be a thick disk star.

\subsection{An over-inflated brown dwarf?}

\begin{figure*}
  \begin{center}
    \includegraphics[width=\textwidth]{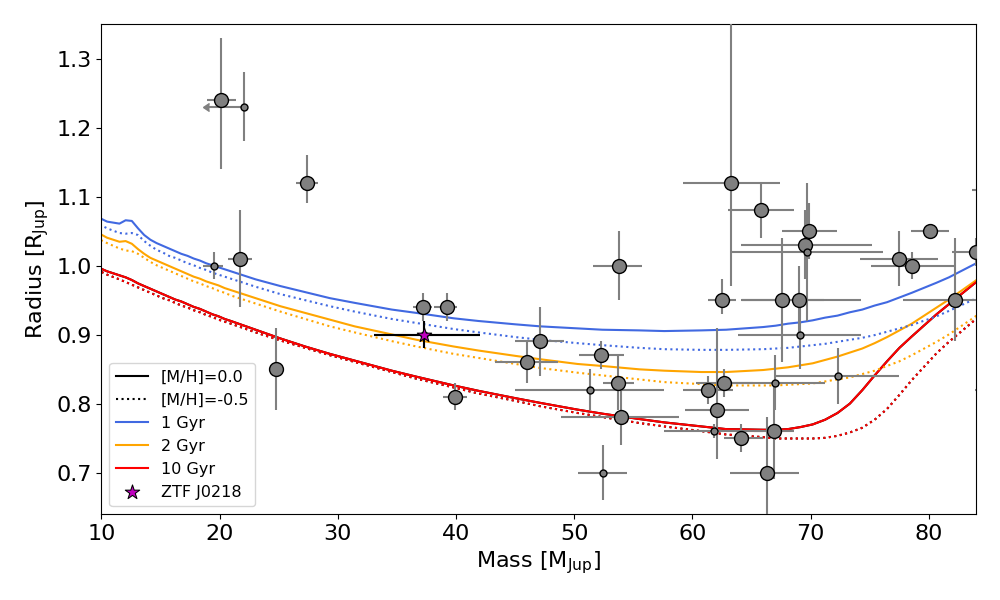}
    \caption{Observed masses and radii of transiting substellar objects with measured ages $>$1\,Gyr (for references see \citealt{Parsons25} and references therein and \citealt{Vowell25}). White dwarf and sdB hosts are shown as small points, while main sequence star hosts are shown as large points. ZTF\,J0218+0711 is highlighted as a magenta star. Also shown are a number of different evolutionary tracks from \citet{Marley21} for ages of 1, 2 and 10 Gyr, both for solar metallicity (solid lines) and [M/H]=-0.5 (dotted lines).}
  \label{fig:BD_radii}
  \end{center}
\end{figure*}

Thanks to the additional constraint provided by the H$\alpha$ emission line, we were able to simultaneously measure both the mass and radius of the brown dwarf from the light curve, allowing us to compare the measured radius to theoretical models. This is dependent on the white dwarf following a standard mass-radius relationship. However, as previously noted, we expect the effects of the white dwarf's magnetic field to have a negligible impact on its measured parameters \citep{Hardy23}, and measured white dwarf parameters show good agreement with theoretical models \citep{Parsons17}. We show the measured brown dwarf parameters in Figure~\ref{fig:BD_radii}, compared to other transiting brown dwarfs and evolutionary models.

As a member of the galactic thick disk it is likely that ZTF\,J0218+0711 is roughly 10\,Gyr old (see Section~\ref{sec:cv} for a more detailed discussion of the total age of the system). For this age and the measured mass of the brown dwarf, we would expect a radius of roughly 0.83\,R$_\mathrm{Jup}$. Our measured radius of $0.90$\,R$_\mathrm{Jup}$ is therefore 8 per cent larger than expected and is more consistent with a roughly 2\,Gyr old object. Over-inflation is often seen for brown dwarfs, both in binaries with white dwarfs \citep[e.g.]{Amaro23,Casewell20,Casewell24,French24,Parsons25} as well as main sequence stars \citep[e.g.][]{Siverd12,Acton21,Adams25}, though the exact cause of this is still unclear and many other brown dwarfs do have radii consistent with theoretical models \citep[e.g.][]{Johnson11,Parsons17b}.

ZTF\,J0218+0711 likely has a low metallicity, given that it is a member of the thick disk. Low metallicity reduces the atmospheric opacity, resulting in more efficient cooling and hence smaller radii \citep{Marley21}, therefore the likely low metallicity makes the radius discrepancy worse, although the effect of metallicity is minor at the mass of the brown dwarf in ZTF\,J0218+0711 (see Figure~\ref{fig:BD_radii}).

Brown dwarfs in close binaries with white dwarfs can be subject to intense irradiation from the much hotter white dwarf. Irradiation can heat (or prevent the cooling of) brown dwarfs, potentially causing them to appear over-inflated \citep{Komacek17}. Indeed, models that take irradiation into account tend to show much better agreement with measurements than models that do not \citep[e.g.][]{Mukherjee25}. However, in these models the irradiating source is usually a solar-type star and it is still unclear exactly what effect the much higher ultraviolet emission of white dwarfs has on the interiors of brown dwarfs, since this light tends to be absorbed by the higher layers of the brown dwarf's atmosphere \citep{Lothringer20}. The white dwarf in ZTF\,J0218+0711 has a temperature of only 8500\,K (and a luminosity of just 0.0006\,L$_\odot$), hence we do not expect it to significantly irradiate the brown dwarf at present. However, the white dwarf would have been significantly hotter in the past and it has been suggested that the cumulative effect of long term irradiation may eventually result in over-inflation (by slowing down the contraction of the brown dwarf), which may only become noticeable once the white dwarf has already significantly cooled \citep{French24}. While this could potentially be the cause of the over-inflation of the brown dwarf in ZTF\,J0218+0711, we note that the brown dwarf in ZTF\,J0038+2030 appears to have a radius consistent with models \citep{Roestel21} and the system hosts a similarly cool white dwarf ($T_\mathrm{eff} = 10900$\,K). So it is unclear if irradiation is the cause of the over-inflation in ZTF\,J0218+0711.

ZTF\,J0218+0711 may instead be over-inflated due to possessing a thick cloud layer \citep{Burrows11}, although with the current data it is impossible to say whether this is the case. Alternatively, the binary may once have been a cataclysmic variable. If this is the case then the brown dwarf could have originally been much more massive (potentially even a stellar mass object) and driven out of thermal equilibrium. Even after the system detaches the brown dwarf would still be significantly over-inflated, since the thermal timescale of brown dwarfs is extremely long. We will now discuss the possibility of this scenario and the evidence in favour of it.

\subsection{A detached cataclysmic variable?} \label{sec:cv}

ZTF\,J0218+0711 appears to be consistent with the model of \cite{Schreiber23}, in which the system was once a standard non-magnetic period bounce CV, then the emergence of the magnetic field of the white dwarf pushed the two stars apart, creating the detached system we see today. However, the alternative possibility is that ZTF\,J0218+0711 is a pre-CV, that exited the common envelope phase as a white dwarf plus brown dwarf binary and has been evolving to shorter periods via gravitational wave radiation ever since.

To test if we can rule out the pre-CV scenario we attempted to reconstruct the past evolution of the system, including the common envelope phase, assuming that it is a pre-CV and following the method of \citet{Zorotovic22}. Initially we placed no constraints on the total age of the system and used the current system parameters to estimate the possible range of $\alpha_\mathrm{CE}$ (the common envelope efficiency parameter) that can reconstruct the system. Typical values for $\alpha_\mathrm{CE}$ are found to be around 0.2-0.4 for white dwarfs with M dwarf and substellar companions \citep[e.g.][]{Zorotovic10,Zorotovic22}, although several white dwarf plus brown dwarf binaries appear to require much larger values, or fine-tuned evolution \citep{Lagos21,Parsons25}. We found that we could reconstruct the evolution of ZTF\,J0218+0711 for any value of $\alpha_\mathrm{CE}$ for both solar and sub-solar metallicity. The ease with which we can reconstruct the system as a pre-CV implies that ZTF\,J0218+0711 could well be a pre-CV system. However, when we forced the total age of the system to be consistent with the Galactic thick disk ($\sim$10\,Gyr) we were unable to reconstruct the evolution of the system regardless of the value of $\alpha_\mathrm{CE}$. No model could reach such an old total age while still creating a white dwarf with the observed mass and temperature. This is because the current white dwarf parameters imply a cooling age of only $1.1 \pm 0.6$\,Gyr \citep{Bedard20} meaning that the progenitor lifetime needs to be close to 9\,Gyr. Hence, it would be a low mass star, which would create a much lower mass white dwarf than is observed, even assuming a perfectly efficient common envelope phase.

While the thick disk is an old Galactic population, there is an observed spread of ages, with the youngest population around 6.6\,Gyr old \citep{Lian25}. However, even assuming a total age of just 6.6\,Gyr we still struggled to reconstruct the evolution of the system, requiring it to have a very low common envelope efficiency of $\alpha_\mathrm{CE} < 0.02$, which would make it an extreme outlier. Assuming a more typical value of $\alpha_\mathrm{CE} = 0.2$ we were unable to create any system older than a total age of 4.5\,Gyr, far younger than even the youngest populations of the thick disk. Therefore, if ZTF\,J0218+0711 is confirmed to be a thick disk star then this would mean that the system must be a detached CV, rather than a pre-CV. If the system was a CV in the past then its current parameters do not accurately reflect the original post-common envelope stellar masses or the current cooling age of the white dwarf. It is therefore possible that the true cooling age of the white dwarf is far longer than the $\sim$1\,Gyr implied by its current temperature, bringing the total age of the system into agreement with the thick disk. 

This scenario could also help explain the over-inflation of the brown dwarf. In order to test this, we computed simulations of the evolution of a detached CV using MESA \citep[][version 24.03.1]{Jermyn23}. A white dwarf with mass 0.688\,{\MSUN} was placed in a binary with a 0.55\,{\MSUN} donor in an initial orbital period of 0.55 days and evolved until contact and into the cataclysmic variable stage. During the CV stage, mass transfer rates were determined using the prescription of \citet{Ritter88}. Angular momentum loss was included as described in \citet{Knigge11}; above the gap magnetic wind braking was reduced from the prescription of \citet{Rappaport83} by a factor of 0.66. Below the gap, a residual wind braking was simulated by including additional angular momentum loss of 1.47 times the gravitational radiation rate. As in \citet{Knigge11} we accounted for the inflation due to Roche distortion, and the known tendency of models to under predict the radii of main-sequence stars by adding starspots to the surface of the donor, which inflated the radius of the donor by $\sim$10 per cent. The resulting CV track closely resembles the "optimal" track published in \citet{Knigge11} and therefore closely recreates the observed radii of CV donor stars \citep{McAllister19}. Once the donor reached a mass of 0.036\,{\MSUN} we terminated mass transfer and followed the evolution of the donor star.

Assuming that ZTF\,J0218+0711 is a detached CV, we used the current white dwarf temperature to estimate how long ago accretion ceased. White dwarf temperatures in period bounce CVs are typically 10-12 000 K \citep{Pala22}. To reach the current white dwarf temperature would therefore take 0.6 -- 0.8 Gyr \citep{Bedard20} after the cessation of mass transfer. After 0.6--0.8 Gyr the detached donor has a radius of 0.092 -- 0.094\,{\RSUN} - in excellent agreement with the observed radius of the secondary in ZTF\,J0218+0711. The predicted temperature of the donor star is 1100 -- 1200 K, consistent with our non-detection during eclipse in the $z$-band.

It is also worth noting that the system parameters (orbital period, masses, white dwarf temperature and Roche-lobe fill factor) are remarkably consistent with the predictions of \citet{Schreiber23} for a detached period bouncer. ZTF\,J0218+0711 also closely resembles the magnetic white dwarf plus brown dwarf binary SMSS\,J1606-1000 \citep{Kawka21}, which is also a thick disk system and therefore poses the same challenge as ZTF\,J0218+0711 to explain the relatively high white dwarf mass and short cooling age with the large total age of a thick disk system if it is a pre-CV. If the scenario of \citet{Schreiber23} is correct, then ZTF\,J0218+0711 and SMSS\,J1606-1000 may represent the ultimate fate for the majority of CVs. However, a dedicated search for similar systems is required to determine whether a sufficient number exist to account for the significant missing population of predicted period bounce CVs.

\section{Conclusions}

We have found that ZTF\,J0218+0711 is an eclipsing binary consisting of a magnetic white dwarf with a substellar companion in a 102 minute orbit. Follow up high-speed HiPERCAM photometry and phase-resolved X-shooter spectroscopy and FORS2 spectropolarimetry revealed that the white dwarf possesses a 19\,MG magnetic field and has a mass of $0.69\pm0.01$\,{\MSUN} and an effective temperature of $8510\pm130$\,K, while the substellar companion is a $37\pm5$\,{\MJUP} slightly over-inflated brown dwarf. Analysis of the Gaia kinematics imply that ZTF\,J0218+0711 is a high probability thick disk system. However, we were unable to reconstruct the evolution of the system assuming that it is a pre-cataclysmic binary because the relatively short cooling age and high white dwarf mass are at odds with the very long total lifetime required for a thick disk star. We suggest instead that the binary used to be a cataclysmic variable, but became detached as a result of the late emergence of the magnetic field of the white dwarf after the system had already become a period bouncer. This scenario can explain the apparent age discrepancy, since the current temperature of the white dwarf does not accurately reflect its total cooling age if it has spent time accreting from its companion.

It is possible that this evolutionary scenario is the final fate for many, if not most, cataclysmic variables. However, confirmation of this requires the identification of a significant population of similar systems. At present only one other system, SMSS\,J1606-1000, shares similar properties to ZTF\,J0218+0711 and so it remains unclear if the apparently missing population of period bouncers could be hiding as detached magnetic white dwarf plus brown dwarf binaries.

\section*{Acknowledgements}

The results presented in this paper are based on observations collected at the European Southern Observatory under programme IDs 113.D$-$0277 and 114.D$-$0066 and on observations made with the Gran Telescopio Canarias (programme ID GTC119-23B), installed in the Spanish Observatorio del Roque de los Muchachos of the Instituto de Astrof{\'i}sica de Canarias, on the island of La Palma. 

SGP acknowledges support by the Science and Technology Facilities Council (grant ST/B001174/1). ARM acknowledges support from MINECO under the PID2023-148661NB-I00 grant and by the AGAUR/Generalitat de Catalunya grant SGR-386/2021. RMO is funded by INTA through grant PRE-OBSERVATORIO and acknowledges support from project PID2023-146210NB-I00 funded by MICIU/AEI/10.13039/501100011033 and by ERDF/EU. MZ acknowledges support from FONDECYT (grants 1250525 and 1221059). VSD and HiPERCAM are funded by the Science and Technology Facilities Council (grant ST/Z000033/1). MRS thanks for support from FONDECYT (grant No.\,1221059). This project has received funding from the European Research Council under the European Union's Horizon 2020 research and innovation programme (Grant agreement numbers 101002408 - MOS100PC).

%%%%%%%%%%%%%%%%%%%%%%%%%%%%%%%%%%%%%%%%%%%%%%%%%%
\section*{Data Availability}

Raw and reduced X-shooter and FORS2 data are available through the ESO archive. Raw and reduced HiPERCAM data are available through the GTC public archive. 

%%%%%%%%%%%%%%%%%%%% REFERENCES %%%%%%%%%%%%%%%%%%

% The best way to enter references is to use BibTeX:

\bibliographystyle{mnras}
\bibliography{ztf0218} % if your bibtex file is called example.bib

% Alternatively you could enter them by hand, like this:
% This method is tedious and prone to error if you have lots of references
%\begin{thebibliography}{99}
%\bibitem[\protect\citeauthoryear{Author}{2012}]{Author2012}
%Author A.~N., 2013, Journal of Improbable Astronomy, 1, 1
%\bibitem[\protect\citeauthoryear{Others}{2013}]{Others2013}
%Others S., 2012, Journal of Interesting Stuff, 17, 198
%\end{thebibliography}

%%%%%%%%%%%%%%%%%%%%%%%%%%%%%%%%%%%%%%%%%%%%%%%%%%

%%%%%%%%%%%%%%%%% APPENDICES %%%%%%%%%%%%%%%%%%%%%

%\appendix

%\section{Some extra material}

%%%%%%%%%%%%%%%%%%%%%%%%%%%%%%%%%%%%%%%%%%%%%%%%%%

% Don't change these lines
\bsp	% typesetting comment
\label{lastpage}
\end{document}